\documentclass[10pt,conference]{IEEEtran}
\AtBeginDocument{%
  }

\usepackage{subcaption}
\usepackage{times}
\usepackage{latexsym}
\usepackage{graphicx}
\usepackage{amsmath}
\usepackage{amssymb}
\usepackage[T1]{fontenc}
\usepackage{fvextra} % Extended verbatim support
\usepackage{enumitem}
\usepackage[utf8]{inputenc}

\usepackage{microtype}

\usepackage{booktabs}
\usepackage{url}
\begin{document}

%%
%% The "title" command has an optional parameter,
%% allowing the author to define a "short title" to be used in page headers.
% StyleSpeech2: 
\title{Generalized Multilingual Text-to-Speech Generation with Language-Aware Style Adaptation}

\author{
\IEEEauthorblockN{Haowei Lou\IEEEauthorrefmark{1}, Hye-young Paik\IEEEauthorrefmark{1}, Sheng Li\IEEEauthorrefmark{2}, Wen Hu\IEEEauthorrefmark{1}, Lina Yao\IEEEauthorrefmark{1}\IEEEauthorrefmark{3}}
\IEEEauthorblockA{\IEEEauthorrefmark{1}School of Computer Science and Engineering, UNSW Sydney, Kensington 2033, Australia\\
\texttt{\{haowei.lou, h.paik, wen.hu, lina.yao\}@unsw.edu.au}} \\
\IEEEauthorblockA{\IEEEauthorrefmark{2}School of Engineering, Institute of Science Tokyo, Yokohama, Japan\\
\texttt{sheng.li@ieee.com}} \\
\IEEEauthorblockA{\IEEEauthorrefmark{3}CSIRO's Data61, Eveleigh 2015, Australia\\
\texttt{lina.yao@data61.com.au}}
}

\maketitle

\begin{abstract}
Text-to-Speech (TTS) models can generate natural, human-like speech across multiple languages by transforming phonemes into waveforms. However, multilingual TTS remains challenging due to discrepancies in phoneme vocabularies and variations in prosody and speaking style across languages. Existing approaches either train separate models for each language, which achieve high performance at the cost of increased computational resources, or use a unified model for multiple languages that struggles to capture fine-grained, language-specific style variations.
In this work, we propose LanStyleTTS, a non-autoregressive,  language-aware style adaptive TTS framework that standardizes phoneme representations and enables fine-grained, phoneme-level style control across languages. This design supports a unified multilingual TTS model capable of producing accurate and high-quality speech without the need to train language-specific models.
We evaluate LanStyleTTS by integrating it with several state-of-the-art non-autoregressive TTS architectures. Results show consistent performance improvements across different model backbones. Furthermore, we investigate a range of acoustic feature representations, including mel-spectrograms and autoencoder-derived latent features. Our experiments demonstrate that latent encodings can significantly reduce model size and computational cost while preserving high-quality speech generation. The demo page can be found at \url{https://lanstyletts.github.io/demo/}.

\end{abstract}
\section{Introduction}

% Add LLM to tokens, and say the limitation is it cost too-much compuation
Text-to-speech (TTS) models are becoming increasingly popular in modern multimedia applications. By converting text to human-like speech, TTS models power many applications such as virtual assistants, navigation tools, and audiobooks that facilitate seamless human-computer interaction. Recent advances in deep learning and artificial intelligence have significantly accelerated the development of TTS models. These developments are driving TTS models toward greater controllability, expressiveness, and naturalness in speech generation. 

Traditional deep learning-based TTS pipelines, such as Tacotron~\cite{wang2017tacotron, shen2018natural}, follow an autoregressive sequence-to-sequence paradigm that converts text into a sequence of phonemes, which are then transformed into acoustic features such as mel-spectrograms. These spectrograms are subsequently used by a vocoder to reconstruct the final waveform. While effective in producing natural-sounding speech, Tacotron-based models suffer from limitations such as slow inference and word-skipping errors~\cite{ren2019fastspeech}, primarily due to their sequential, autoregressive nature.
To address these issues, non-autoregressive models like FastSpeech~\cite{ren2019fastspeech, ren2020fastspeech} have been proposed. These models adopt Transformer-based architectures~\cite{vaswani2017attention} and parallel inference, which significantly improves inference speed and robustness. However, FastSpeech requires ground-truth phoneme duration annotations for training, making the process data-intensive. Additionally, the accuracy of the duration annotations and prediction module can greatly influence the overall performance of the TTS system.
Later models such as Glow-TTS~\cite{kim2020glow} and VITS~\cite{kim2021conditional} address this limitation by integrating phoneme duration modeling into end-to-end frameworks. These approaches reduce reliance on manually labeled duration data, which in turn improves both training efficiency and generated speech quality. 

Despite these technical advancements, the broader deployment of TTS remains limited by high training costs and poor generalization across languages. Conventional multilingual TTS often require training separate models for each language, a process that is resource-intensive and difficult to scale.
A unified multilingual TTS model would offer substantial gains in efficiency and maintainability. However, this goal remains challenging due to the phoneme mismatches across languages and the presence of language-specific stylistic features, such as tone in tonal languages and stress in non-tonal languages. 

To address phoneme-level variability, recent research has explored the use of the International Phonetic Alphabet (IPA)~\cite{zhang2021revisiting}, which offers a standardized representation of phonemes across languages. While IPA-based methods hold promise for unifying multilingual phoneme tokenization, they still struggle to effectively capture and disentangle language-specific stylistic cues. For instance, tonal distinctions in Chinese carry lexical meaning, whereas in languages like English, stress patterns play an important linguistic role.

Several works have attempted to address these limitations. ZMM-TTS~\cite{gong2024zmm} supports multilingual speech generation but primarily focuses on non-tonal languages and lacks explicit control over tonal variation. CosyVoice~\cite{du2024cosyvoice} and FishSpeech~\cite{liao2024fish} introduce an LLM-guided grapheme-to-phoneme transformation and support multilingual synthesis; however, their reliance on autoregressive decoding limits their efficiency, making them unsuitable for real-time applications. StyleSpeech~\cite{lou2024stylespeech} introduces a style decorator module to enable fine-grained control over tone but is limited to monolingual settings. In parallel, Zhang et al.~\cite{zhang2021revisiting} propose an IPA-based cross-lingual TTS framework~(IPA-TTS) that models phoneme tokens and language-specific style embeddings separately. However, their reliance on a simple additive embedding fusion strategy constrains the model's expressiveness and leads to suboptimal performance.
To the best of our knowledge, there is a lack of research on multilingual, non-autoregressive TTS models that can effectively control fine-grained, phoneme-level stylistic variations across different languages.

In this work, we present \textbf{LanStyleTTS}. A multilingual, non-autoregressive and \textbf{Lan}guage-aware \textbf{Style} adaptive \textbf{TTS} framework that addresses the limitations of previous models by introducing a more effective fusion technique for phoneme-level style control. Specifically, LanStyleTTS leverages the International Phonetic Alphabet (IPA) to standardize phoneme representations across languages and incorporates a novel style adaptation module capable of modeling tonal and prosodic variations in both tonal and non-tonal languages. This design enables efficient and accurate speech generation across multiple languages within a single unified model.

We evaluate the effectiveness of the language-specific style adaptation module in LanStyleTTS across several state-of-the-art non-autoregressive TTS models, including FastSpeech2~\cite{ren2020fastspeech} and VITS~\cite{kim2021conditional}. In all cases, models augmented with our style adaptation module consistently outperform their baseline counterparts. It demonstrates the generalizability and effectiveness of our approach.
In addition, we investigate the impact of different acoustic feature representations on multilingual TTS performance. Specifically, we compare conventional mel-spectrograms with latent encodings derived from an autoencoder. Our experimental results show that latent representations can significantly reduce model size and computational cost while preserving speech quality.
The main contributions of LanStyleTTS are as follows:
\begin{itemize}
    \item We propose a novel style adaptation module that captures language-specific stylistic features and enables fine-grained control of tone and prosody across both tonal and non-tonal languages.
    \item We leverage the IPA phonetic system to unify phoneme representations across languages, improving generalizability in multilingual TTS.
    \item  We validate our framework through extensive experiments across multiple state-of-the-art non-autoregressive TTS backbones and demonstrate improvements in both the naturalness and accuracy of generated speech.
    \item We conduct a systematic comparison of acoustic feature representations and show that latent features significantly reduce model size and inference time while preserving high-quality speech output.
\end{itemize}

\begin{figure*}[ht]
    \centering
    \begin{subfigure}{0.64\textwidth}
        \centering
        \includegraphics[width=\textwidth]{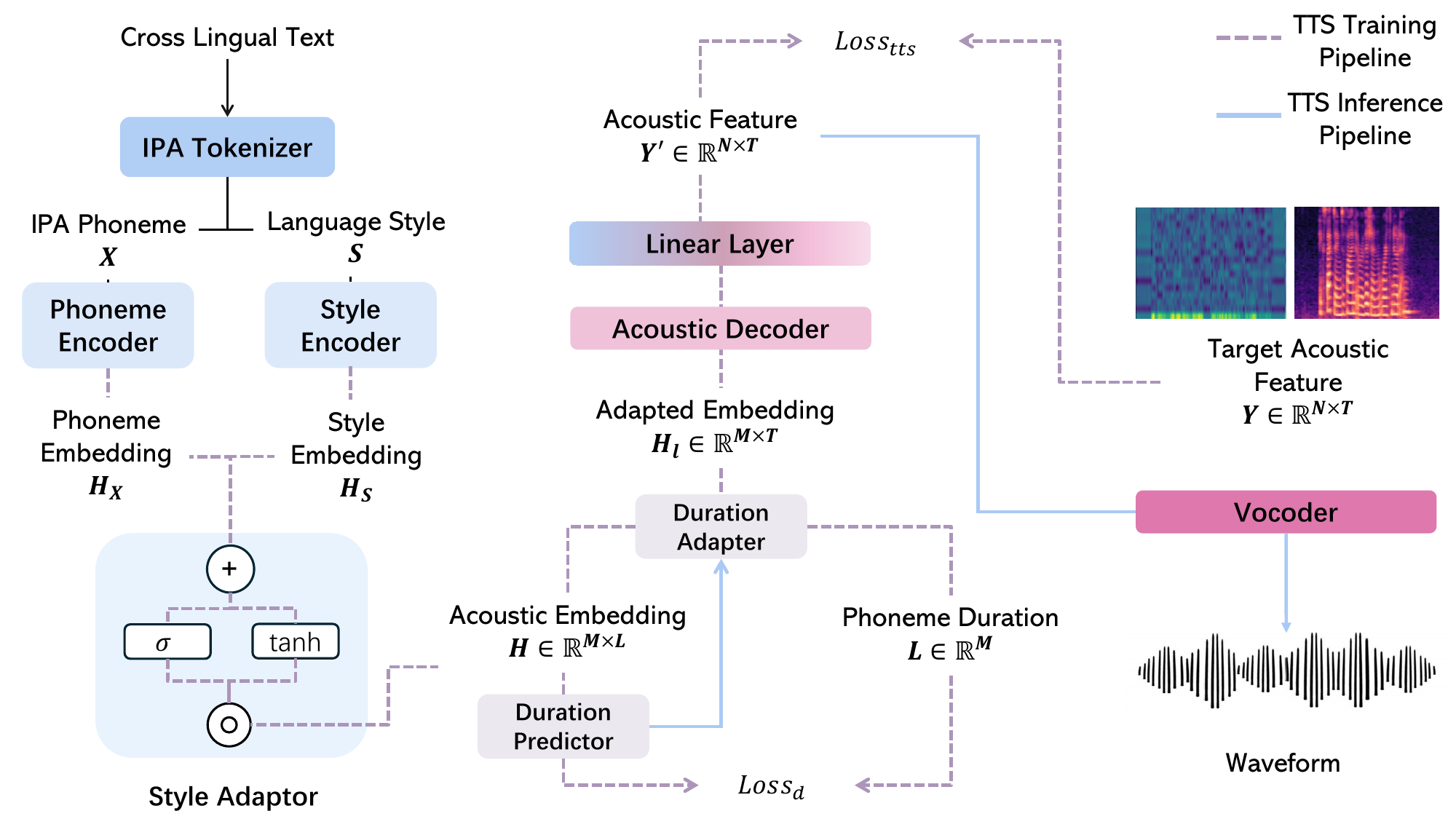}
        \caption{LanStyleTTS-Base}
        \label{fig:stylespeech2}
    \end{subfigure}
    \hfill
    \begin{subfigure}{0.34\textwidth}
        \centering
        \includegraphics[width=\textwidth]{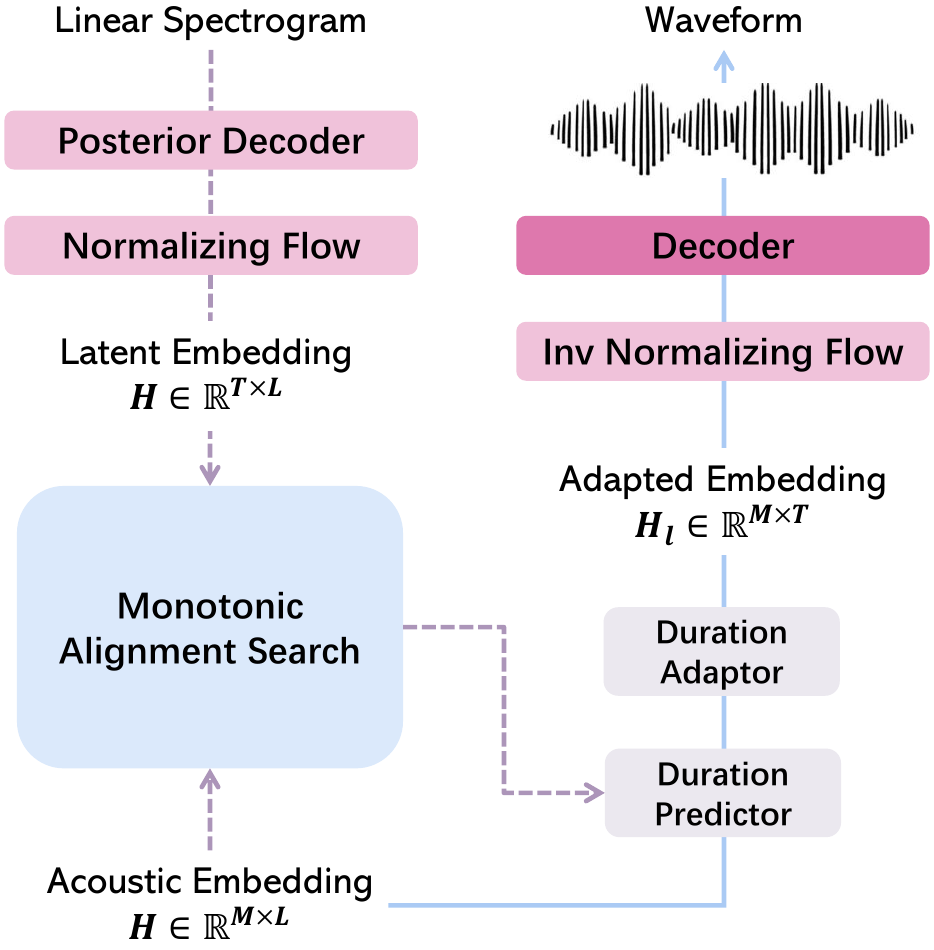}
        \caption{LanStyleTTS-VITS}
        \label{fig:stylespeech2_vits}
    \end{subfigure}
    \caption{LanStyleTTS Architecture Overview}
\end{figure*}

\section{Method}

To accommodate both FastSpeech-based and VITS-based architectures, we introduce two variants: \textbf{LanStyleTTS-base}, which follows the standard non-autoregressive, FastSpeech backbone, and \textbf{LanStyleTTS-VITS}, which integrates the VITS backbone as the acoustic decoder for end-to-end waveform generation.
In the conventional non-autoregressive TTS framework, exemplified by FastSpeech2, the speech generation process begins by tokenizing the input text into a sequence of phoneme tokens, denoted as $X \in \mathbb{R}^L$. $L$ represents the length of the token sequence.  The model then transforms $X$ into an acoustic feature sequence $Y \in \mathbb{R}^{N \times T}$, where $N$ is the feature dimension and $T$ is the length of the output sequence.  A separate vocoder subsequently converts $Y$ into waveform.

LanStyleTTS has \textbf{five} main components: a text tokenizer that converts written text into phoneme tokens $X$ and style tokens $S$; encoders that create phoneme and style embeddings $H_X$ and $H_S$; a style adaptation module that combines the phoneme and style embeddings into a single acoustic embedding $H$ and adjusts the duration of each frame to create the adaptive embedding $H_l$; an acoustic decoder and linear layer that turn the adaptive embedding into an acoustic feature sequence $Y$; and a vocoder that converts the acoustic features into a waveform. Detailed explanations of each module are provided in the following sections.

LanStyleTTS-VITS, on the other hand, builds on the variational inference framework of VITS. It integrates phoneme, style, and pitch information into an end-to-end model that maps phoneme embeddings directly to waveform. There is no need for a separate vocoder. Detailed explanation will be presented in Section~\ref{sec:vits}.

\subsection{Text Tokenizer}\label{sec:text_token}
% Text Tokenizer plays a crucial role as the first processing step by converting text (also known as graphemes) into a tokenized sequence that the TTS model can process. Phonemes are closer to the acoustic properties of speech, making them more favorable for the TTS process. This conversion process is commonly referred to as grapheme-to-phoneme (G2P) conversion.

In this research, we explore two tokenization schemes. First, we use character-based tokenization, where text is segmented at the character level using English alphabet characters. Second, we apply phoneme-based tokenization, where text is split into words and each word is converted into IPA phonemes. More details on these tokenization methods will be discussed in the following subsections. An example of the tokenization process is shown in Table \ref{tab:phoneme_comparison}.

% use the IPA-based tokenizer to convert both English and Chinese graphemes into phonemes. Additionally, we employ a standard alphabet-based tokenization as part of an ablation study to evaluate the effectiveness of using the IPA system.

\begin{table}[t]
    \centering
    \resizebox{\linewidth}{!}{
    \begin{tabular}{l|l@{\hskip 1pt}l@{\hskip 3pt}l|l@{\hskip 1pt}l@{\hskip 3pt}l} % Adjust column spacing
        \toprule
        & \multicolumn{3}{c}{\textbf{English}} & \multicolumn{3}{c}{\textbf{Chinese}} \\
        & \textbf{Good} & & \textbf{Day} & \textbf{ni3} & & \textbf{hao3} \\
        \midrule
        Alphabet & \texttt{|good} & | & \texttt{day |} & \texttt{|ni} & | & \texttt{hao |} \\
        \midrule
        Phoneme  & \texttt{|G UH1 D} & | & \texttt{D EY1|} & \texttt{|n i3} & | & \texttt{h ao3|} \\
        IPA      & \texttt{|g u d} & | & \texttt{d ei |} & \texttt{|n i } & | & \texttt{x au |} \\
        Style    & \texttt{00 1 0} & \texttt{0} & \texttt{0 1 0} & \texttt{00 3} &\texttt{0}& \texttt{0 3 0} \\
        \bottomrule
    \end{tabular}
    }
    \caption{Language-Aware Phoneme Tokenization}
    \label{tab:phoneme_comparison}
\end{table}

\subsubsection{Alphabet Characters}
Alphabet is a simple and efficient character-based tokenization scheme. Its key advantages are its flexibility in character selection and ease of processing. However, a major limitation is that characters are not phonemes. In most linguistic studies, language-specific tonal information is associated with phonemes rather than individual characters. This results in character-based encoding lacking the mapping for its style.

In this encoding scheme, we construct an alphabet dictionary consisting of 27 characters: letters A-Z and a silent character \( |\), which serves as a word separator.
We first split a sentence into a sequence of words. Each character in a word is then treated as an individual token. For non-character-based words, such as numbers, the number is first converted to its word representation before applying the same tokenization method. A silent token \( | \) is used to separate consecutive words. 
For Chinese text, each Chinese character is first converted into its corresponding Pinyin representation. Each Pinyin are then tokenized into alphabet sequences using the same method as for English.
% \begin{verbatim}
% good day => | g o o d | d a y |
% ni3 hao3 => | n i | h a o |
% \end{verbatim}

% \begin{table}[t]
%     \centering
%     \renewcommand{\arraystretch}{1}
%     \resizebox{\linewidth}{!}{
%     \begin{tabular}{l|l|l}
%         \hline
%         & English: Good Day & Pinyin: ni3 hao3 \\
%         \hline
%         Characters & \texttt{| g o o d | d a y |} & \texttt{| n i | h a o |} \\
%         \hline
%     \end{tabular}
%     }
%     \caption{IPA Phonemes Tokenization}
%     \label{tab:phoneme_comparison}
% \end{table}

\subsubsection{IPA Phonemes}
We select the union of Chinese and English-related IPA phonemes, resulting in a set of 81 phonemes.
Additionally, due to the richness of the IPA system, we can incorporate language-specific stylized elements of phonemes. These include tone variations in Chinese and vowel stress in English.
In total, our IPA dictionary consists of 89 elements: 81 phonemes, 5 tone markers for Chinese, and 3 stress markers for English. 

The tokenization process for English involves first converting each word into English phonemes using the CMU Pronouncing Dictionary~\cite{cmu_pronouncing_dict}.
The stress marker for vowels is removed from the phoneme sequence to form the style element sequence, and the English phonemes are then mapped to their corresponding IPA phonemes.

For Chinese, the tokenization process starts by transforming each character into its Pinyin representation using the Python pypinyin library. Each Pinyin word is divided into initials and finals. The tonal marker in the final is removed to form the style element sequence. Finally, the initials and finals are converted into their corresponding IPA phonemes.

% \begin{verbatim}
% English: Good Day
% Phoneme: | G UH1 D | D EY1 |
% IPA:     | g u   d | d ei  |
% Style:   0 0 1   0 0 0 1   0

% Pinyin:  ni3 hao3
% Phoneme: | n i3 | h ao3 |
% IPA:     | n i  | x au  |
% Style:   0 0 3  0 0 3   0
% \end{verbatim}

\subsection{Token Encoder}\label{sec:token_encoder}
Once we collect the tokenized phoneme $X$ and style $S$ sequence $X,S \in \mathbb{R}^L$. The next step is to encode these tokens into the sequence of embedding, denoted as $X_H,S_H \in \mathbb{R}^{M \times L}$, where $M$ is the embedding dimension. 

Then, we apply positional encoding transformations to $X_H$ and $S_H$ to ensure the model retains positional information for the sequential data. 
Subsequently, we employ several layers of Feed-Forward Transformer(FFT) block from FastSpeech~\cite{ren2019fastspeech} to process the embeddings. These FFT blocks are designed to efficiently handle sequential data by capturing both local and global dependencies within the token sequences. Each FFT block incorporates multi-head self-attention to capture token-wise relationships within the sequences. The standard two-layer dense network in the Transformer~\cite{vaswani2017attention} is replaced with two 1DCNNs. This modification is aimed at improving the model's ability to capture local dependencies between adjacent phoneme embeddings.

\subsection{Style Adapter}\label{sec:style_adapter}
In linguistic studies, there are many stylistic linguistic elements that can be used to modify a phoneme or control the tone, pitch, or stress of speech. We collectively refer to these elements as Style. In many tonal languages, such as Chinese, Vietnamese, and Thai, tone is used to distinguish lexical or grammatical meanings within a sentence \cite{kisseberth2007tone}. This creates a challenge for TTS models that rely purely on phonetic tokens, as they cannot precisely articulate the nuances of speech. As a result, such models have limited controllability and, consequently, reduced performance.

The Style Adapter is specifically designed to learn and adapt to these stylistic elements. Given the phoneme embedding $H_X$ and the style token embedding $H_S$ from Section~\ref{sec:token_encoder}, we first fuse them by adding them together and acoustic embedding $H$ is computed as:
\[H' = H_X + H_S\]
\[H = \mathrm{tanh}(H’) \odot \mathrm{sigmoid}(H')\]
where   $\odot$ denotes element-wise multiplication. +

Due to the discrepancy between the length of the acoustic feature sequence $T$ and the phoneme token sequence $L$, as well as the language-specific variations in speaking patterns, we employ a duration predictor and a duration adapter to control the duration of each phoneme. Since no existing library supports calculating the duration of IPA phonemes, we trained an IPA phoneme aligner following the approach described in \cite{lou2024aligner} and used force alignment to obtain the duration of each phoneme, $l \in \mathbb{R}^{L}$, where $\sum l = T$.

We first train a duration predictor $\mathbb{DP}$ to estimate the duration of each phoneme based on its embedding. The predicted duration is defined as $l' = \mathbb{DP}(H)$, and the duration loss $Loss_{\text{d}}$  is calculated as: 

\begin{equation} Loss_{\text{d}} = \left| \log(l) - \log(l') \right| \end{equation}
During the training stage, the ground truth duration label $L$ is used to adjust the duration of each frame in the acoustic embedding, producing the adaptive embedding $H_l$.

During inference, we use the learned duration predictor to estimate the length of each acoustic frame and adjust their duration to ensure smooth and natural speech generation.

\subsection{Acoustic Decoder}
The duration-adapted embeddings $H_l \in \mathbb{R}^{M \times T}$ are processed through an acoustic decoder to generate the target output. The acoustic decoder consists of multiple FFT blocks, followed by a linear projection layer that projects the embeddings to match the dimensions of the target acoustic feature $Y$, producing the final output $Y' \in \mathbb{R}^{N \times T}$. 

In this research, we selected two types of acoustic features as the target for the TTS model. The first is the Mel-Spectrogram, a widely used representation of speech that captures its time-frequency structure. The second is the latent feature, obtained by training an autoencoder following the settings from \cite{lou2024latentspeech}. The audio is encoded into a latent feature using the trained encoder, and this latent feature is treated as the target acoustic feature for the TTS model. The TTS loss $Loss_{\text{tts}}$ and final loss is computed as is calculated as: 
\begin{equation} Loss_{\text{tts}} = \left| Y - Y' \right|^2 \end{equation}
\begin{equation}  Loss = Loss_{\text{d}} + Loss_{\text{tts}}\end{equation}
\subsection{Vocoder}
The vocoder is responsible for transforming the acoustic feature $Y$ into speech audio. 

For the Mel-Spectrogram, we utilize the pretrained WaveGlow model~\cite{prenger2019waveglow} from NVIDIA to convert the spectrogram into audio. WaveGlow is a flow-based generative model that combines the benefits of WaveNet and Glow architectures. It uses an invertible $1 \times 1$ convolution and affine coupling layers to model the distribution of raw audio waveforms, conditioned on the input Mel-Spectrogram. By sampling from a Gaussian noise distribution and applying the learned transformations, WaveGlow produces natural-sounding audio with efficient inference.

For the latent feature, we employ our custom-trained Autoencoder's decoder to reconstruct audio from the latent representations. In practice, both types of acoustic features can generate clear and intelligible audio. A more detailed analysis of their performance will be presented in the discussion section.

\subsection{VITS Variant}\label{sec:vits}
The LanStyleTTS-VITS variant follows the process as LanStyleTTS-Base for generating text tokens (Section~\ref{sec:text_token}), encoding text token to embedding using the text encoder (Section~\ref{sec:token_encoder}), and adapting language-specific style using the style adapter (Section~\ref{sec:style_adapter}). In the subsequent stage, the VITS framework~\cite{kim2021conditional} is adopted to enable end-to-end duration modeling and waveform generation.

VITS functions as a conditional variational autoencoder (VAE), optimizing the evidence lower bound (ELBO) of the log-likelihood of the speech signal conditioned on the acoustic embedding $H \in \mathbb{R}^{M \times L}$ from Section~\ref{sec:style_adapter}. A posteior encoder and invertible normalizing flow~\cite{rezende2015variational} is employed to transform the linear spectrogram derived from waveform into a latent embedding $Z \in \mathbb{R}^{M \times T}$. During training, Monotonic Alignment Search (MAS)~\cite{kim2020glow} is applied to align the acoustic embedding $H$ with the latent representation $Z$ in a fully differentiable manner. Additionally, a discriminator is introduced within an adversarial training framework to further improve the naturalness and quality of the generated speech. Figure~\ref{fig:stylespeech2_vits} present an overview of the LanStyleTTS-VITS Variant.

\begin{table*}[h!]
\centering
\caption{WER and MOS Evaluation Split by Chinese (CH) and English (EN) Compared to Baseline Models; Best Scores in \textbf{Bold}}
\label{tab:seperate}
% \resizebox{\linewidth}{!}{
\begin{tabular}{lcccc}
\hline
\textbf{Method} & \multicolumn{2}{c}{\textbf{WER (\%)~(\(\downarrow\))}} & \multicolumn{2}{c}{\textbf{MOS~(\(\uparrow\))}} \\
\hline
 & \textbf{CH} & \textbf{EN}  & \textbf{CH} & \textbf{EN}  \\
\hline
FastSpeech2~\cite{ren2019fastspeech} & 25.00 $\pm$ 22.99 & 12.53 $\pm$ 22.83 & 1.87 $\pm$ 0.16 & 2.02 $\pm$ 0.24 \\
StyleSpeech~\cite{lou2024stylespeech} & 31.20 $\pm$ 15.60 & N/A & 1.48 $\pm$ 0.15 & N/A \\
VITS~\cite{kim2021conditional}  & 37.32 $\pm$ 18.26 & 10.67 $\pm$ 15.23 & 2.77 $\pm$ 0.16 & 4.52 $\pm$ 0.35 \\
IPA-TTS~\cite{zhang2021revisiting} & 8.26 $\pm$ 12.31 & 11.46 $\pm$ 19.29 & 1.95 $\pm$ 0.18 & 2.01 $\pm$ 0.31 \\
\hline
Ours \\
\hline
LanStyleTTS Base   & 7.82 $\pm$ 10.60 & \textbf{8.24 $\pm$ 11.70} & 2.43 $\pm$ 0.18 & 2.23 $\pm$ 0.28 \\
LanStyleTTS VITS  & \textbf{5.14 $\pm$ 7.01} & 9.52 $\pm$ 15.30  & \textbf{3.93 $\pm$ 0.54} & \textbf{4.63 $\pm$ 0.14} \\
% Style Adaptation \\
% \hline
% StyleSpeech2 (w s)   & 7.82 $\pm$ 10.60 & 8.24 $\pm$ 11.70 & 2.92 $\pm$ 0.48 & 2.98 $\pm$ 0.40 \\
% StyleSpeech2 (w/o s)  & 25.98 $\pm$ 21.46 & 11.72 $\pm$ 15.32  & 2.49 $\pm$ 0.39 & 2.73 $\pm$ 0.44  \\
% \hline
% Duration Label \\
% \hline
% StyleSpeech2 (w l)  & 5.66 $\pm$ 7.21 & 8.69 $\pm$ 13.82  &  3.32 $\pm$ 0.51 & 2.93 $\pm$ 0.40  \\
% StyleSpeech2 (w/o l)    & 7.82 $\pm$ 10.60 & 8.24 $\pm$ 11.70  & 2.92 $\pm$ 0.48 & 2.98 $\pm$ 0.40  \\
% \hline
% Token Type \\
% \hline
% StyleSpeech2 (Alphabet)  & 37.75 $\pm$ 19.89 & 35.47 $\pm$ 34.63  & 2.35 $\pm$ 0.36 & 2.55 $\pm$ 0.42    \\
% StyleSpeech2 (IPA)  & \textbf{7.82 $\pm$ 10.60} & \textbf{8.24 $\pm$ 11.70}  & \textbf{2.92 $\pm$ 0.48} & \textbf{2.98 $\pm$ 0.40} \\
% StyleSpeech2 + VITS (Alphabet) & 27.61 $\pm$ 15.38 & 13.58 $\pm$ 19.98  & 2.14 $\pm$ 0.13 & 3.56 $\pm$ 0.56  \\
% StyleSpeech2 +  VITS (IPA) & \textbf{5.14 $\pm$ 7.01} & \textbf{9.52 $\pm$ 15.30}  & \textbf{2.98 $\pm$ 0.73} & \textbf{3.88 $\pm$ 0.58} \\
% \hline 
% Vocoder \\
% \hline
% MelSpec + WaveGlow & 8.05 $\pm$ 16.23 & 10.48 $\pm$ 15.04   & \textbf{3.12 $\pm$ 0.52} & \textbf{3.27 $\pm$ 0.51} \\
% Latent + AE Decoder  &\textbf{ 7.82 $\pm$ 10.60} & \textbf{8.24 $\pm$ 11.70}  &2.92 $\pm$ 0.48 & 2.98 $\pm$ 0.40   \\
\hline
Ground Truth & 1.95 $\pm$ 4.11 & 6.16 $\pm$ 10.20   & 4.90 $\pm$ 0.22 & 4.92 $\pm$ 0.14   \\
\hline
\end{tabular}
% }%
\end{table*}

\begin{table*}[h!]
\centering
\caption{Ablation Studies on Impact of Style Adaptation Module,  Phoneme Token Types, and Acoustic Features}
\label{tab:ablation}
% \resizebox{\linewidth}{!}{
\begin{tabular}{lcccc}
\hline
\textbf{Method} & \multicolumn{2}{c}{\textbf{WER (\%)~(\(\downarrow\))}} & \multicolumn{2}{c}{\textbf{MOS~(\(\uparrow\))}} \\
\hline
 & \textbf{CH} & \textbf{EN}  & \textbf{CH} & \textbf{EN}  \\
\hline
\textbf{Style Adaptation} \\
\hline
LanStyleTTS Base (w/o s)  & 25.98 $\pm$ 21.46 & 11.72 $\pm$ 15.32  & 1.96 $\pm$ 0.12 & 2.07 $\pm$ 0.28  \\
LanStyleTTS Base (w s)   & \textbf{7.82 $\pm$ 10.60} & \textbf{8.24 $\pm$ 11.70} & \textbf{2.43 $\pm$ 0.18} & \textbf{2.23 $\pm$ 0.28} \\
\hline\
LanStyleTTS VITS (w/o s)  & 37.32 $\pm$ 18.26 & 10.67 $\pm$ 15.23 & 2.77 $\pm$ 0.16 & 4.52 $\pm$ 0.35  \\
LanStyleTTS VITS (w s)   & \textbf{5.14 $\pm$ 7.01} & \textbf{9.52 $\pm$ 15.30}  & \textbf{3.93 $\pm$ 0.54} & \textbf{4.63 $\pm$ 0.14} \\
\hline
\textbf{Token Types} \\
\hline
LanStyleTTS Base (Alpha)  & 37.75 $\pm$ 19.89 & 35.47 $\pm$ 34.63  & 1.82 $\pm$ 0.32 & 1.81 $\pm$ 0.28    \\
LanStyleTTS Base (IPA Only)  & 25.98 $\pm$ 21.46 & 11.72 $\pm$ 15.32  & 1.96 $\pm$ 0.12 & 2.07 $\pm$ 0.28  \\
LanStyleTTS Base (IPA + Style)  & \textbf{7.82 $\pm$ 10.60} & \textbf{8.24 $\pm$ 11.70}  & \textbf{2.43 $\pm$ 0.18} & \textbf{2.23 $\pm$ 0.28} \\
\hline
LanStyleTTS VITS (Alpha) & 27.61 $\pm$ 15.38 & 13.58 $\pm$ 19.98  & 3.01 $\pm$ 0.28 & 4.58 $\pm$ 0.30  \\
LanStyleTTS VITS (IPA Only)  & 37.32 $\pm$ 18.26 & 10.67 $\pm$ 15.23 & 2.77 $\pm$ 0.16 & 4.52 $\pm$ 0.35  \\
LanStyleTTS VITS (IPA + Style) & \textbf{5.14 $\pm$ 7.01} & \textbf{9.52 $\pm$ 15.30}  & \textbf{3.93 $\pm$ 0.54} & \textbf{4.63 $\pm$ 0.14} \\
\hline
\textbf{Acoustic Feature \& Vocoder} \\
\hline
MelSpec + WaveGlow & 8.05 $\pm$ 16.23 & 10.48 $\pm$ 15.04   & 2.37 $\pm$ 0.49 & \textbf{2.47 $\pm$ 0.43} \\
Latent + AE Decoder  &\textbf{ 7.82 $\pm$ 10.60} & \textbf{8.24 $\pm$ 11.70}  & \textbf{2.43 $\pm$ 0.18} & 2.23 $\pm$ 0.28  \\
\hline
\end{tabular}

\end{table*}

\section{Experiment}\label{sec:experiment}
To evaluate the model's generalizability across different languages, we selected English and Chinese for evaluation, as they are among the most widely spoken languages globally and represent both tonal (Chinese) and non-tonal (English) languages. We used the Baker dataset~\cite{BakerDataset2020} and LJSpeech dataset~\cite{ljspeech17} for training and evaluation. Specifically, the Baker dataset contains 10,000 Chinese speech samples, while the LJSpeech dataset comprises 13,100 English speech samples. All speech data were resampled to 48 kHz for consistency.

We selected 9,000 Chinese and 9,000 English training speech samples to ensure balanced training. In total, there are approximately 27 hours of speech audio in the training set. An additional 1,000 speech samples from each language were reserved for testing the model's performance. The batch size was set to 32, and the model was trained for 500 epochs (approximately 280,000 iterations) on a single NVIDIA V100 GPU. The training process takes approximately 44 hours to complete. The phoneme embedding size was set to 256, and the learning rate followed the inverse square-root schedule with a 4,000-iteration warm-up to ensure stable convergence.

\subsubsection{Baselines}
To evaluate the performance of our proposed LanStyleTTS framework, we compare it against several non-autoregressive baseline models in a multilingual setting. For FastSpeech2\cite{ren2020fastspeech}, we train the model using latent acoustic features as supervision targets.  For StyleSpeech\cite{lou2024stylespeech}, we use the official open-source implementation. However, as it only supports Chinese, our evaluation is restricted to the Chinese subset of our dataset. For VITS\cite{kim2021conditional}, we train the model on phoneme sequences under the same multilingual configuration. This model serves as a strong end-to-end baseline, capable of generating waveforms directly without relying on an external vocoder. Lastly, we reimplement IPA-TTS~\cite{zhang2021revisiting} based on the StyleSpeech architecture and integrate the additive feature fusion mechanism introduced in the original design.

\subsubsection{Metrics}
We evaluated our method using both subjective and objective metrics to assess the TTS model from two key perspectives: accuracy and naturalness. For the objective metric, we employed Word Error Rate (WER). First, we generated speech using the TTS model and then transcribed it with OpenAI's Whisper~\cite{radford2023robust}. Transcriptions were compared to the original text, with a lower WER indicating better generation accuracy.

For the subjective metric, we invited five participants fluent in both English and Chinese to participate in the evaluation. The participants listened to various speech samples generated by different models and provided overall ratings for the clarity and naturalness of each sample. All samples were shuffled to prevent bias or leakage of model information, ensuring a fair evaluation process. The Mean Opinion Score (MOS) was then calculated as the subjective metric, primarily reflecting the naturalness of the generated speech.
Thus, WER was used to measure accuracy, while MOS focused on evaluating the naturalness of the TTS output.

% 58940.204 english, 38205.16 chinese
\subsection{Ablation Study}
We compare the performance of the TTS model trained using alphabet-based phoneme tokens and IPA tokens to assess their impact on the generalizability and quality of the TTS model. Furthermore, we analyze the effect of the style adaptation module by training and evaluating the TTS model with and without this component, examining its influence on prosody, naturalness, and overall performance. Finally, we compare the performance and resource requirements of the TTS model trained and inferred with different acoustic features. Specifically, we evaluate models trained on Mel-Spectrograms and latent features to determine which representation offers a better balance between quality audio generation and computational efficiency.

\section{Discussion}
The main goal of our study is to address the following three key research questions (\textbf{RQs}):
\begin{itemize}
\item \textbf{RQ1}:  Can the proposed multilingual style adaptation method effectively capture language-specific speaking styles and enhance the performance of multilingual TTS models?
\item \textbf{RQ2}:  Which tokenization method provides a more generalized and effective approach for generating high-quality speech in multilingual TTS models?
\item \textbf{RQ3}: Which acoustic feature, Mel-Spectrogram or latent feature, is more efficient and effective for generating high-quality speech samples in TTS models?
\end{itemize} 

Table~\ref{tab:seperate} presents an overall performance comparison across various methods, while Table~\ref{tab:ablation} details the results of the ablation studies. To establish an upper bound for TTS model performance, we include ground truth speech as a reference. The evaluation follows the protocol described in Section~\ref{sec:experiment}.

Table~\ref{tab:seperate} shows that our proposed method consistently outperforms existing baselines in both WER and MOS across languages. Specifically, for the base model, the WER is reduced from 8.26\% to 7.82\% in Chinese and from 11.46\% to 8.24\% in English. The MOS also improves significantly, rising from 1.95 to 2.43 for Chinese and from 2.01 to 2.23 for English. With further enhancements, the WER for Chinese is reduced to 5.14\%, while the MOS scores improve to 3.93 and 4.63 for Chinese and English, respectively. Although the VITS variant shows a slightly higher English WER (9.52\%) compared to the base approach (8.24\%), it still substantially outperforms the VITS baseline, which records a WER of 10.67\% in English. These results underscore the effectiveness of our proposed approach in a multilingual TTS setting. 

Table~\ref{tab:ablation} presents ablation studies evaluating the impact of the style adaptation module, phoneme token types, and acoustic features on multilingual TTS performance. Models equipped with the style adaptation module consistently outperform their counterparts without it across both Chinese and English. For instance, the WER of the base model drops significantly from 25.98\% to 7.82\% in Chinese and from 11.72\% to 8.24\% in English, while MOS improves from 1.96 to 2.43 and from 2.07 to 2.23, respectively. The VITS variant exhibits a similar trend, with the Chinese WER reduced from 37.32\% to 5.14\% and the MOS increased from 2.77 to 3.93, demonstrating the effectiveness of the proposed style adaptation mechanism.
In the second section, we compare IPA-based tokenization with traditional alphabet-based tokenization. The IPA + Style variant achieves dramatically better results: for the base model, the WER improves from 37.75\% to 7.82\% (Chinese) and from 35.47\% to 8.24\% (English), while MOS also increases substantially. This confirms that IPA tokenization offers better multilingual generalization and phonetic consistency.
The third section compares two types of acoustic features. The model using latent features with an autoencoder-based decoder outperforms the Mel-spectrogram + WaveGlow baseline in 3 out of 4 metrics, achieving lower WERs and higher MOS in most cases. These results highlight the superior representation capability of latent features in capturing expressive and natural-sounding speech. 
A more detailed analysis of these findings will be presented in the subsequent section.

% Overall, StyleSpeech2 outperforms the baseline models in terms of both accuracy and naturalness. For the choice of acoustic features, traditional Mel-Spectrograms with a Vocoder demonstrate slightly better performance compared to the proposed latent features. This performance difference is primarily driven by the results in English.

% When examining language-specific performance in Table~\ref{tab:seperate}, the AE-Latent model achieves superior results in Chinese, outperforming all other models in this language. However, its relatively low performance in English contributes to a lower overall score. The reasons behind this disparity are further discussed in Section~\ref{sec:limpact}.

\begin{figure*}[ht!]
    \centering
    % Alphabet embeddings visualization
    \begin{subfigure}{0.32\textwidth}
        \centering
        \includegraphics[width=\textwidth]{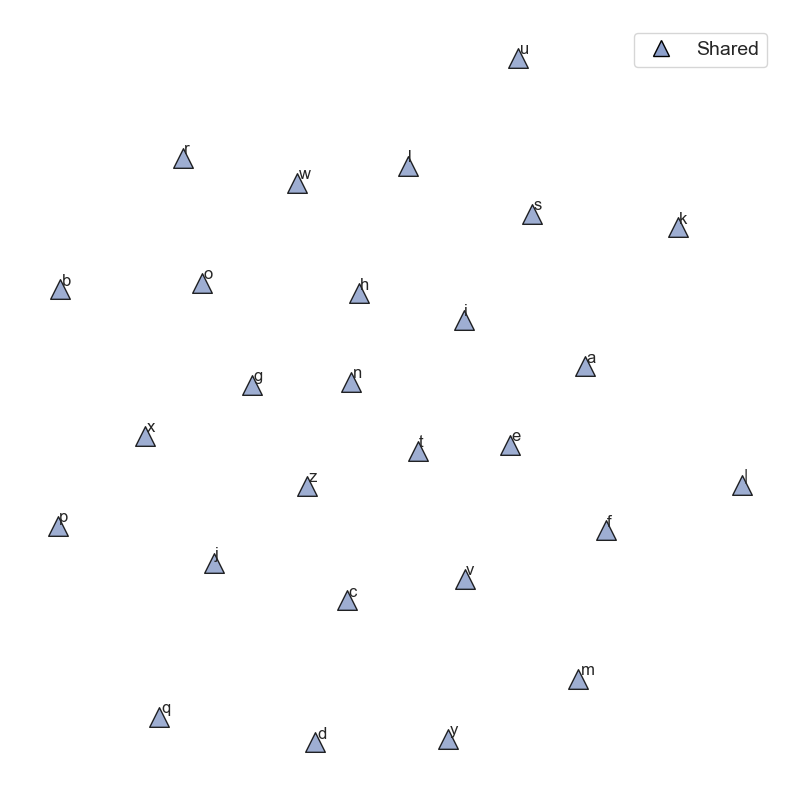} 
        \caption{Alphabet}
        \label{fig:alphabet_embeddings}
    \end{subfigure}
    \hfill
    % IPA embeddings visualization
    \begin{subfigure}{0.32\textwidth}
        \centering
        \includegraphics[width=\textwidth]{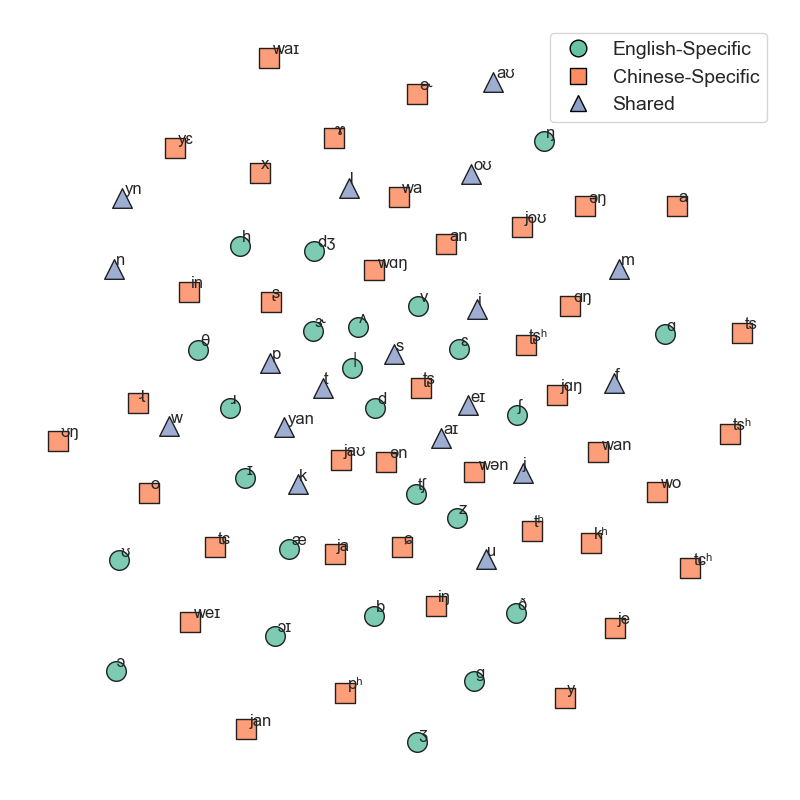} 
        \caption{IPA without language-specific style}
        \label{fig:ipa_embeddings}
    \end{subfigure}
    \hfill
    % IPA with language-specific style embeddings visualization
    \begin{subfigure}{0.32\textwidth}
        \centering
        \includegraphics[width=\textwidth]{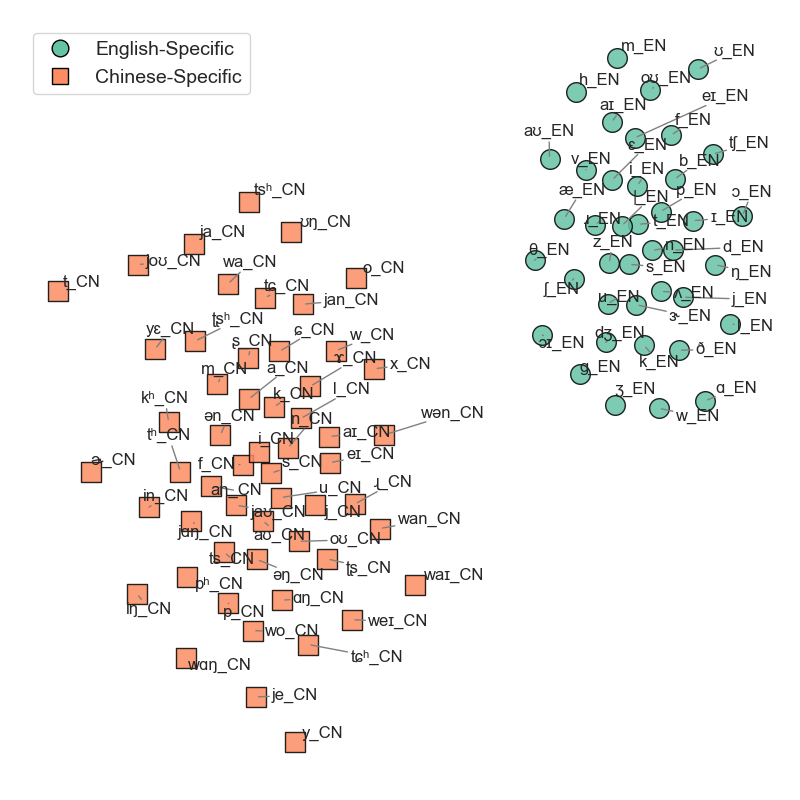} 
        \caption{IPA with language-specific style}
        \label{fig:ipa_style_embeddings}
    \end{subfigure}

    \caption{Phoneme Embedding Visualization. (a) Alphabet embeddings. (b) Phoneme embeddings without language-specific style. (c) Phoneme embeddings with language-specific style..}
    \label{fig:embedding_visualizations}
\end{figure*}
\subsection{Style Adaption Module - RQ1}
The style adaptation module plays a critical role in enhancing the performance of our proposed method. As shown in Table~\ref{tab:seperate} and further detailed in Table~\ref{tab:ablation}. Incorporating the module (w/ s) in TTS models leads to a substantial improvement in both intelligibility (WER) and naturalness (MOS) across different model variants.

For the base model, the inclusion of the style adaptation module reduces the WER from 25.98\% to 7.82\% in Chinese and from 11.72\% to 8.24\% in English. MOS also increases from 1.96 to 2.43 (Chinese) and from 2.07 to 2.23 (English). The performance gain is especially prominent for Chinese, where tonal variation plays a key role in speech intelligibility. These results demonstrate that the module effectively captures and applies language-specific style and enables clearer and more expressive TTS generation.

Notably, the benefits of the style adaptation module extend beyond a specific model architecture. In the VITS-variant, which uses a different vocoder and acoustic modeling strategy, the style-adaption module also yields significant improvements. For example, Chinese WER drops dramatically from 37.32\% to 5.14\%, and MOS rises from 2.77 to 3.93. English performance also improves, with MOS increasing from 4.52 to 4.63, despite the relatively high baseline. This indicates that the style adaptation mechanism is robust and transferable. It can enhance TTS model's performance even in models with more complex or adversarial training setups like VITS.

Overall, these findings confirm that our proposed style adaptation module enables fine-grained control over phoneme-level style features and contributes consistently to higher-quality, more natural speech generation, regardless of the underlying model architecture or language.

\subsection{Impact of Phoneme Tokenization - RQ2}
Phoneme tokenization, which converts written text into units suitable for speech generation, is a crucial first step in any TTS pipeline. The choice of tokenization scheme has a direct impact on the model's ability to generate accurate and natural speech. While language-specific tokenizers have shown success in monolingual settings, developing a robust, generalized multilingual tokenization strategy remains an ongoing challenge.

In our experiments, IPA-based phoneme tokenization significantly outperformed character-level, alphabet-based tokenization. In our experiments, IPA-based phoneme tokenization significantly outperformed character-level, alphabet-based tokenization. Specifically, it led to a 30\% reduction in WER for Chinese, 27\% for English, along with MOS improvements of 33\% and 23\%, respectively. These gains highlight the superiority of IPA for both generation quality and multilingual generalizability.

The improvement primarily stems from the phonetic precision of IPA-based phonemes. Unlike alphabet-based schemes, which are rooted in abstract orthographic conventions, IPA provides a direct and consistent representation of pronunciation. For example, the letter ``k'' is silent in \textit{know} but voiced in \textit{kitty}, introducing inconsistencies that models must learn to resolve. Similarly, the letter ``a'' is pronounced as /ei/ in \textit{fake} but as /æ/ in \textit{fat}, depending on the word. IPA eliminates such ambiguity by encoding only audible phonemes, making it easier for the model to align phoneme tokens with the acoustic signal and reducing the overall learning burden.

The core issue lies in the fact that IPA phonemes may share the same symbol form across languages but differ in their phonetic realizations and prosodic patterns. For example, the IPA symbol /i/ appears in both English (e.g., \textit{see}) and Chinese (e.g., \textit{xī}), but its acoustic realization differs across the two languages. In English, /i/ is typically diphthongized into [i] or [ij], especially in certain dialects, whereas in Mandarin Chinese, it is produced as a pure, monophthongal high front vowel. Without language-specific adaptation, a multilingual TTS model may conflate these variations, leading to unnatural or inconsistent speech outputs across languages.

These observations underscore the importance of incorporating a \textbf{language-specific style adaptation} alongside IPA tokenization. While IPA improves low-level phonetic alignment, style adaptation captures higher-level, language-dependent variations such as tone and stress. Together, they enable the model to generate speech that is not only phonetically accurate but also stylistically appropriate across different languages.

To further support this, we visualize the learned phoneme embedding spaces for different tokenization methods in Figure~\ref{fig:embedding_visualizations}. We use T-SNE for dimensionality reduction. English and Chinese-specific phonemes are colored in green and orange, respectively, while shared phonemes are labeled in blue. As shown in Figures~\ref{fig:alphabet_embeddings} and \ref{fig:ipa_embeddings}, both alphabet- and IPA-based embeddings form fairly balanced distributions. However, Figure~\ref{fig:ipa_style_embeddings}, which includes embeddings enriched with language-specific style adaption, reveals clearly disjoint clusters between languages after style adaptio.
This separation indicates that our model not only leverages IPA to accurately encode phonetic content, but also effectively captures language-specific stylistic variations within the embedding space.

\subsection{Impact of Acoustic Feature - RQ3}

Most existing TTS models, including FastSpeech 1\&2~\cite{ren2019fastspeech,ren2020fastspeech}, Glow-TTS~\cite{kim2020glow}, and StyleSpeech~\cite{lou2024stylespeech}, adopt a two-stage pipeline: the TTS model first generates a Mel-Spectrogram as a target ouput, which is then converted into waveform audio using a neural vocoder. While this approach has proven effective, it introduces several limitations, prompting researchers to explore alternative acoustic representations.

In our evaluations, we found that speech generated via WaveGlow~\cite{prenger2019waveglow} often deviated from the original speaker's voice characteristics. For example, instead of preserving the mature female voice present in the training data, the generated speech frequently resembled that of a young boy. This highlights a major drawback of the two-step approach: the dependency on a vocoder can compromise speaker fidelity. Achieving high-quality results often requires vocoder-specific fine-tuning or retraining, which is time-consuming, inefficient, and impractical for large-scale or multilingual applications.

To address these limitations, we explore an alternative approach in which the TTS model directly predicts a latent acoustic representation. In this setup, the latent feature replaces the Mel-Spectrogram as the model's target output, removing the reliance on vocoder-specific artifacts. Notably, even in models like VITS, which generate speech directly, the feature matching space can still be interpreted as a form of latent embedding. VITS encodes the linear spectrogram into a latent feature, and alignment occurs within that learned space. However, since VITS does not involve explicitly supervised feature matching, we exclude it from direct comparison in this study.

Our results, shown in Table~\ref{tab:seperate}, indicate that under equivalent conditions, models utilizing latent features outperform their Mel-Spectrogram counterparts, particularly in terms of WER and MOS for Chinese. This suggests that latent representations can offer not only better intelligibility but also improved speaker representation in multilingual scenarios.
However, subjective evaluations reveal nuanced trade-offs. Based on listener feedback, speech generated using Mel-Spectrograms is often perceived as more fluid and natural. While the AE-decoder demonstrates more accurate tonal variation, especially in tonal languages like Chinese. But its output is sometimes described as choppy or fragmented. This could be due to the discretized nature of machine-learned latent features, which may introduce discontinuities in the reconstructed audio. In contrast, the Mel-Spectrogram, which more closely resembles the original acoustic signal, provides smoother temporal transitions that align well with human auditory expectations.

These findings underscore the importance of selecting appropriate acoustic features not only for intelligibility and speaker accuracy but also for perceived naturalness. Latent features offer promising advantages, but their integration may require further refinement to fully match the perceptual quality of traditional Mel-Spectrogram-based speech generation.

We compared the resource requirements for generating audio using different acoustic feature representations, as summarized in Table~\ref{tab:resouce_comparision}. In terms of inference speed, decoding with latent features is significantly faster than vocoder-based approaches. Specifically, generating an 8-second audio clip using the Mel-Spectrogram approach takes approximately 7 seconds, while the latent feature approach completes the same task in just 0.4 seconds. The VITS variant requires around 0.8 seconds.

In addition to faster inference, models using latent features are also more lightweight. The model size for the Mel-Spectrogram approach is approximately 320M parameters, compared to just 61M for the latent feature model and 43M for the VITS variant. These substantial reductions in both inference time and model size make the latent feature approach particularly well-suited for resource-constrained environments.

Overall, these advantages position latent features as a compelling alternative to traditional Mel-spectrograms, especially in applications where computational efficiency and deployment scalability are critical.

\begin{table}[t!]
\centering
\caption{Comparison of Inference Time and Parameter Size for Different Vocoder to generate an 8s audio}
\label{tab:resouce_comparision}
\resizebox{\linewidth}{!}{
\begin{tabular}{lcc}
\hline
\textbf{Vocoder} & \textbf{Inference (ms)} & \textbf{\# Parameter (M)} \\
\hline
\textbf{LanStyleTTS}            & 315  & 60.46 \\
\textbf{WaveGlow}        & 6,622  & 268.09 \\
\hline
\textbf{LanStyleTTS}            & 288  & 60.48 \\
\textbf{AE Decoder}              & 142  & 0.86  \\
\hline
\textbf{LanStyleTTS-VITS}            & 810  & 42.52 \\
\hline
\end{tabular}
}
\label{tab:component_comparison}
\end{table}
% average 21.5 item /s, 

\subsection{Other Finding and Future Directions}\label{sec:limpact}
% Silent phonemes, denoted as "$|$," have a significant impact on the performance of TTS models, both during training and inference stages. This effect appears to be more pronounced when using AE latent encodings compared to Mel-spectrograms.

% English words often contain varying combinations of phonemes, while in Chinese, the length variation between phonemes is much smaller. This results in a rhythmic discontinuity, as the training dataset typically uses fixed silent intervals between a fixed number of phonemes. As a result, the model tends to generate speech on a word-by-word basis. In contrast, English does not face such issues. One potential solution to this challenge is incorporating phrases, rather than individual words, during model training. We plan to explore this approach in our future research.

% Another challenge arises during inference when the model's predicted durations (denoted as $l$ labels) do not align well with the ground truth durations. However, when we used ground truth durations during inference, we observed a significant improvement in results, with a greater than 4\% improvement in generation quality for both English and Chinese. While this improvement underscores the importance of accurate duration prediction, relying on ground truth durations is impractical in real-world scenarios.
Our experiments reveal that the VITS variant consistently outperforms the base model in terms of MOS, despite both approaches demonstrating comparable performance in intelligibility and tonal accuracy. According to listener feedback, while the base model, regardless of whether it uses Mel-Spectrogram or latent features, achieves strong tonal variation and phonetic intelligibility.  However, the generated speech often contains undesired noise and artifacts, which negatively impact perceived naturalness and lead to lower subjective ratings.

In contrast, the VITS variant produces cleaner, noise-free speech without such distortions, resulting in consistently higher MOS scores. This indicates that, while the base model benefits from phoneme-level style adaptation and efficient latent representations, it is currently limited by the quality of the latent features derived from the autoencoder during decoding.

This insight highlights a key direction for future work. In our ongoing research, we plan to improve the latent feature learning mechanism within the base framework to enhance robustness and reduce audio artifacts. By addressing this limitation, we aim to further advance the naturalness and overall quality of the base approach while maintaining its efficiency and scalability.

\section{Conclusion}
In conclusion, we propose LanStyleTTS, a generalized multilingual TTS model that integrates a language-aware style adaptation module and IPA-based tokenization to produce natural and intelligible speech across languages. Our experiments and ablation studies demonstrate that both components contribute significantly to performance improvements, with the style adaptation module effectively capturing language-specific speaking styles in both English and Chinese. Moreover, LanStyleTTS is compatible with both traditional two-stage pipelines and end-to-end frameworks such as VITS, and consistently delivers promising improvements in intelligibility (WER) and naturalness (MOS) across different settings.

Additionally, we explore the impact of different acoustic feature on efficiency. Models using latent features decoded through an AE-based framework offer a substantial reduction in inference time and parameter size, making them more suitable for resource-constrained scenarios. While this approach incurs a slight drop in MOS compared to Mel-spectrogram-based systems, it strikes a strong balance between efficiency and quality, positioning LanStyleTTS as a practical and scalable solution for multilingual speech generation.

%%
%% The next two lines define the bibliography style to be used, and
%% the bibliography file.
\bibliographystyle{ACM-Reference-Format}
\bibliography{custom}

%%
%% If your work has an appendix, this is the place to put it.
\appendix

\end{document}